\documentclass{article}
\usepackage{spconf,amsmath,graphicx}
\usepackage{booktabs}
\usepackage{amssymb}
\usepackage{bm}
\usepackage{multirow}
\usepackage{arydshln}
\usepackage{cite}
\usepackage{url}
\usepackage{setspace}

\newcommand{\vardbtilde}[1]{\tilde{\raisebox{0pt}[0.9\height]{$\tilde{#1}$}}}

\makeatletter
\def\adl@drawiv#1#2#3{%
        \hskip.5\tabcolsep
        \xleaders#3{#2.5\@tempdimb #1{1}#2.5\@tempdimb}%
                #2\z@ plus1fil minus1fil\relax
        \hskip.5\tabcolsep}
\newcommand{\cdashlinelr}[1]{%
  \noalign{\vskip\aboverulesep
           \global\let\@dashdrawstore\adl@draw
           \global\let\adl@draw\adl@drawiv}
  \cdashline{#1}
  \noalign{\global\let\adl@draw\@dashdrawstore
           \vskip\belowrulesep}}
\makeatother

\title{Hierarchical Conditional End-to-End ASR with\\CTC and Multi-Granular Subword Units}
\name{
    Yosuke Higuchi,
    Keita Karube,
    Tetsuji Ogawa,
    Tetsunori Kobayashi
}

\address{
    Department of Communications and Computer Engineering, Waseda University, Tokyo, Japan
 }

\begin{document}
\ninept
\maketitle
\setlength{\abovedisplayskip}{4pt}
\setlength{\belowdisplayskip}{4pt}
\begin{abstract}
\vspace{-1mm}
In end-to-end automatic speech recognition (ASR),
a model is expected to implicitly learn representations suitable for recognizing a word-level sequence.
However, the huge abstraction gap between input acoustic signals and output linguistic tokens makes it challenging for a model to learn the representations.
In this work,
to promote the word-level representation learning in end-to-end ASR,
we propose a \textit{hierarchical conditional} model that is based on connectionist temporal classification (CTC).
Our model is trained by auxiliary CTC losses applied to intermediate layers, where
the vocabulary size of each target subword sequence is gradually increased as the layer becomes close to the word-level output.
Here, we make each level of sequence prediction explicitly conditioned on the previous sequences predicted at lower levels.
With the proposed approach,
we expect the proposed model to learn the word-level representations effectively by
exploiting a hierarchy of linguistic structures.
Experimental results on LibriSpeech-\{100h, 960h\} and TEDLIUM2 demonstrate that the proposed model improves over
a standard CTC-based model and other competitive models from prior work.
We further analyze the results to confirm
the effectiveness of the intended representation learning with our model.
\end{abstract}
\vspace{-1mm}
\vspace{-2mm}
\section{Introduction}
\vspace{-2mm}
\label{sec:intro}
End-to-end automatic speech recognition (ASR) aims to model direct speech-to-text conversion~\cite{graves2014towards, chorowski2015attention, chan2016listen}, 
which substantially simplifies the training and inference processes without external knowledge (e.g., a pronunciation lexicon).
With well-established sequence-to-sequence modeling techniques~\cite{graves2006connectionist,graves2012sequence, sutskever2014sequence, bahdanau2014neural} 
and more sophisticated neural network architectures~\cite{dong2018speech,gulati2020conformer,majumdar2021citrinet}, 
end-to-end ASR models have shown promising performance on various benchmarks~\cite{chiu2018state, luscher2019rwth, karita2019comparative}.

Contrary to carefully designed feature extraction in the traditional pipeline framework, 
end-to-end models are generally expected to implicitly learn representations suitable for solving a specific task.
For example,
the learned representations have been shown to represent 
shape features for image classification~\cite{zeiler2014visualizing} and 
syntactic structures for language modeling~\cite{peters2018deep}.
However, in ASR,
it can be more challenging for an end-to-end model to learn representations automatically.
Having no access to segmentation or alignment information,
end-to-end ASR models are required to predict word-level linguistic tokens from
frame-level acoustic signals.
This input-output gap in the level of abstraction makes it difficult to optimize end-to-end ASR,
unless a large amount of data or a strong language model is accessible during training or inference~\cite{zhang2020pushing, irie2019language}.

To promote word-level representation learning in end-to-end ASR,
we believe that a model should be trained to gradually increase the abstraction level of linguistic information,
as it has long been considered reasonable for recognizing speech
(i.e., speech $\rightarrow$ phonemes $\rightarrow$ words $\rightarrow$ text)~\cite{jelinek1976continuous}.
By exploiting lower levels of abstractions to conditionally compose the higher-level linguistic information,
an end-to-end ASR model should be able to handle the sparsity problem of words~\cite{soltau2016neural} and
extract effective representations.

\begin{figure}[t]
    \centering
    \includegraphics[width=0.95\columnwidth]{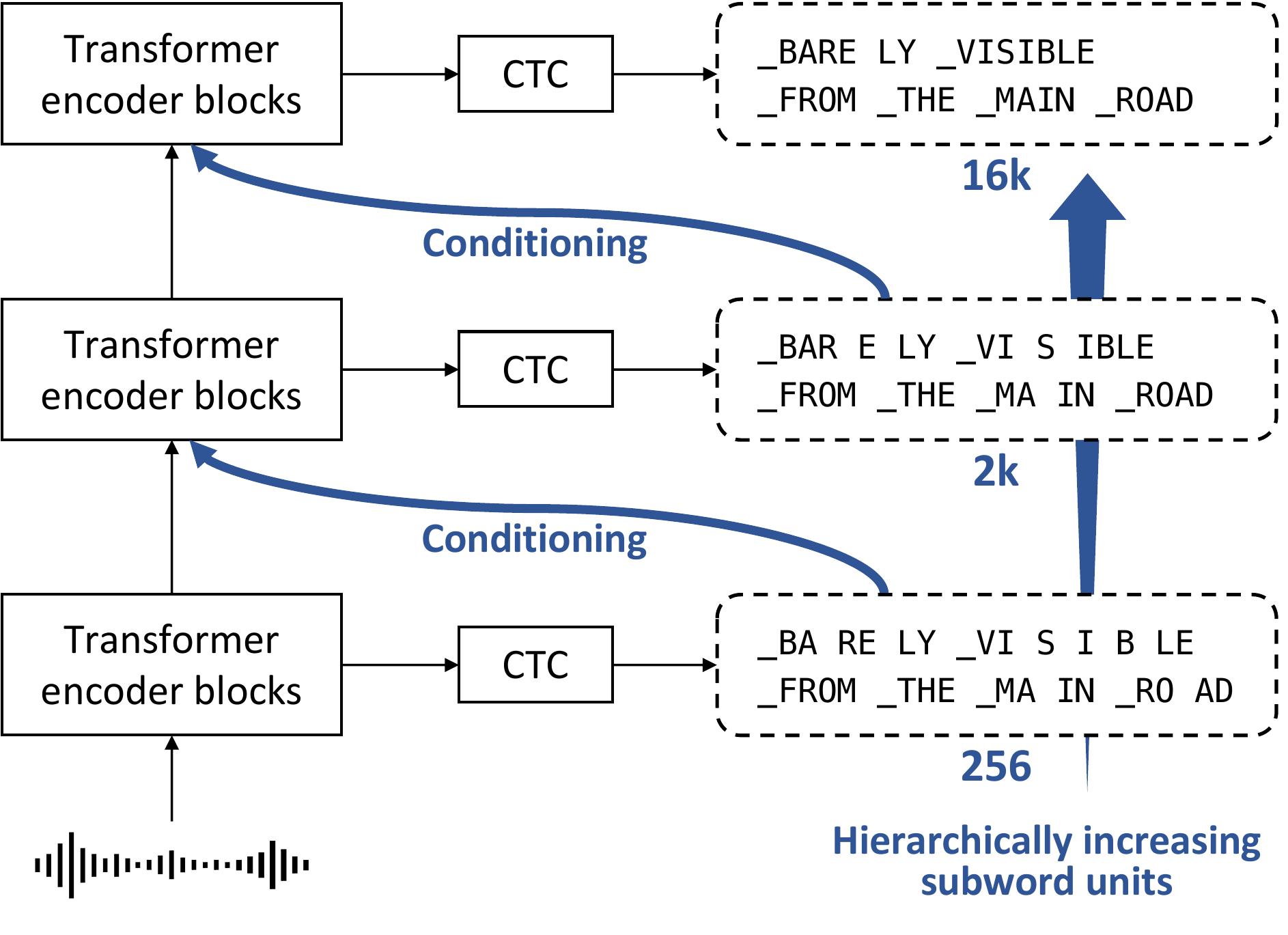}
    \hspace{-3mm}
    \vspace{-4mm}
    \caption{Proposed hierarchical conditional model of end-to-end ASR.}
    \label{fig:proposed}
    \vspace{-6mm}
\end{figure}

To achieve such progressive representation learning for ASR,
we propose \textit{hierarchical conditional} modeling of end-to-end ASR (Figure~\ref{fig:proposed}).
Our model consists of multiple connectionist temporal classification (CTC)~\cite{graves2006connectionist} losses
hierarchically applied to the intermediate and last layers, inspired by previous studies~\cite{fernandez2007sequence,rao2017multi,toshniwal2017multitask,sanabria2018hierarchical,krishna2018hierarchical,tjandra2020deja,lee2021intermediate}.
Each loss calculation targets sequences with a different granularity of linguistic information:
sequences with lower abstraction levels are predicted from the intermediate layers, and 
a word-level sequence is predicted from the last layer.
Specifically, we focus on subwords (n-gram characters) and
increase the vocabulary size to word-level
as the model layer becomes close to the output (e.g., 256 $\rightarrow$ 2k $\rightarrow$ 16k).
In addition to this hierarchical structure,
we design the model to predict each sequence at an abstraction level by
explicitly conditioning on the previously predicted sequences at lower levels,
which is crucial for maintaining subwords attributed to composing the higher-level sequence.
The proposed model should capture a hierarchy of linguistic structures and
yield representations suitable for modeling words.

The key contributions of this work are summarized as follows.
1) We show that the proposed approach enables a CTC-based system to learn accurate word-level ASR,
mitigating the data-sparsity issue by gradually increasing the abstraction level of intermediate predictions.
2) Based on experiments conducted on LibriSpeech and TEDLIUM2, we demonstrate the effectiveness of our model independently of variations in the amount of data and speaking styles. All the implementations are made publicly available on our ESPnet fork ({\footnotesize \url{https://github.com/YosukeHiguchi/espnet/tree/hierctc}}).
3) We carefully compare our model with other CTC-based models and further analyze the results,
which provides in-depth insights into the advantage of the proposed modeling.


\section{Hierarchical Conditional End-to-End ASR}

\label{sec:hierasr}

\subsection{Baseline architecture of end-to-end ASR}
End-to-end ASR is formulated as a sequence-mapping problem between a $T$-length input sequence $X \!=\! (\bm{\mathrm{x}}_t \in \mathbb{R}^D| t\!=\!1,\dots,T)$ and 
$L$-length output sequence $Y \!=\! ( y_l \in \mathcal{V} | l\!=\!1,\dots,L )$. 
Here, $\bm{\mathrm{x}}_t$ is a $D$-dimensional acoustic feature at frame $t$, 
$y_l$ is an output token at position $l$, and $\mathcal{V}$ is a vocabulary. 
As a baseline, we focus on a Transformer-based model~\cite{vaswani2017attention} optimized by CTC~\cite{graves2006connectionist} with intermediate loss calculation~\cite{tjandra2020deja, lee2021intermediate}.

\noindent\textbf{Transformer encoder:}
For encoding an audio sequence $X$ into latent representations, we construct the Transformer encoder~\cite{vaswani2017attention}
consisting of a stack of $E$ self-attention layers.
The $i$-th layer outputs a sequence of $d_{\mathsf{model}}$-dimensional latent representations $X^{(i)}\!=\!(\bm{\mathrm{x}}_t^{(i)}\in\mathbb{R}^{d_{\mathsf{model}}} | t=1,...,T) $ as
\begin{align}
    \tilde{X}^{(i)} &= X^{(i-1)} + \mathrm{SelfAttention}(X^{(i-1)}),\label{eq:tf_enc1} \\
    X^{(i)} &= \tilde{X}^{(i)} + \mathrm{FeedForward}(\tilde{X}^{(i)}),\label{eq:tf_enc2}
\end{align}
where $i \! \in \! \{1,...,E\}$, 
and $X^{(0)}$ is obtained by adding positional encodings to $X$. 
In Eqs.~\eqref{eq:tf_enc1} and~\eqref{eq:tf_enc2}, layer normalization is applied to each input of the self-attention mechanism $\mathrm{SelfAttention(\cdot)}$ and feedforward network $\mathrm{FeedForward(\cdot)}$.
We also train a model with the Conformer encoder~\cite{gulati2020conformer}, 
which introduces a convolution neural network (CNN) into the Transformer encoder, 
i.e., a convolution module is added between Eqs.~\eqref{eq:tf_enc1} and~\eqref{eq:tf_enc2}.

\noindent\textbf{Connectionist temporal classification:}
CTC~\cite{graves2006connectionist} optimizes the model to predict a monotonic alignment between the encoded input $X^{(E)}$ and output $Y$.
To align the sequences in frame-level, 
the output sequence $Y$ is augmented with a unique blank token $\epsilon$, 
which results in a latent token sequence $Z=(z_t \in \mathcal{V} \cup \{\epsilon\}| t=1,\dots,T)$.
On the basis of the conditional independence assumption per token-frame prediction, 
CTC models the conditional probability $P_{\mathsf{ctc}} (Y | X^{(E)})$ by marginalizing over latent token sequences as
\begin{equation}
    \label{eq:p_ctc}
    P_{\mathsf{ctc}} (Y | X^{(E)}) \approx \sum_{Z \in \mathcal{B}^{-1} (Y)} \prod_{t=1}^{T} P (z_t | X^{(E)}), 
\end{equation}
where $\mathcal{B}^{-1}(Y)$ returns all possible latent sequences compatible with $Y$.
The CTC loss is defined as the negative log-likelihood of Eq.~(\ref{eq:p_ctc}): 
\begin{equation}
    \mathcal{L}_{\mathsf{ctc}}(Y | X^{(E)}) = -\log P_{\mathsf{ctc}}(Y|X^{(E)}). 
    \label{eq:ctc_loss}
\end{equation}

\noindent\textbf{Intermediate CTC:} In addition to the standard CTC loss calculated from the model output, 
auxiliary CTC losses can be iteratively applied to intermediate layers~\cite{tjandra2020deja,lee2021intermediate}.
Such intermediate losses effectively regularize the model training and lead to improved ASR performance.
We consider training the model with a total of $K$ CTC losses applied to the output and intermediate layers:
\begin{equation}
    \mathcal{L}_{\mathsf{sc\text{-}ctc}} = \frac{1}{K} \left\{ \mathcal{L}_{\mathsf{ctc}}(Y | X^{(E)}) + \sum_{k=1}^{K-1} \mathcal{L}_{\mathsf{ctc}}(Y | X^{(\lfloor \frac{kE}{K} \rfloor)}) \right\},
    \label{eq:interctc_loss}
\end{equation}
where $1\!<\!K\!\le E$, and
we equally distribute the weight across the losses~\cite{lee2021layer}.
In Eq.~\eqref{eq:interctc_loss},
we adopt the self-conditioning mechanism~\cite{nozaki2021relaxing},
which improves a CTC-based model by relaxing the conditional independence assumption.
For the intermediate layer, from which a CTC loss is calculated, 
we modify Eq.~\eqref{eq:tf_enc2} as
\begin{align}
    \vardbtilde{X}^{(i)} &= \tilde{X}^{(i)} + \mathrm{FeedForward}(\tilde{X}^{(i)}), \label{eq:tf_enc2_modified}\\
    X^{(i)} &= \vardbtilde{X}^{(i)} + \mathrm{Linear}(A^{(i)}), \label{eq:tf_enc3}
\end{align}
where $i\!\in\!\{\lfloor kE / K \rfloor\}_{k=1}^{K-1}$, and $A^{(i)}\!=\!\mathrm{softmax}(\vardbtilde{X}^{(i)})$ is a sequence of the posterior distributions w.r.t latent tokens computed by CTC.

\vspace{-1mm}
\subsection{Subword segmentation}
\vspace{-1.5mm}
For tokenizing target sequences, 
subword segmentation is a widely used approach for alleviating the out-of-vocabulary problem~\cite{sennrich2016neural},
where words in a sentence are split into subword units (or n-gram characters).
In the general algorithm for building a subword vocabulary,
pairs of subword units are repeatedly merged on the basis of the frequency appearing in a text corpus.
The iteration stops when the vocabulary reaches an arbitrary size.

We adopt subwords for tokenizing ASR transcriptions.
As opposed to characters, 
subwords can provide the model with shorter output sequences, thus 
reduce the difficulty of modeling the dependency between outputs.
This can be especially important for CTC-based modeling with the conditional independence assumption.
However, it should be noted that increasing the subword vocabulary size makes a sequence close to word-level and potentially lead to the data-sparsity problem~\cite{soltau2016neural}.



\begin{table*}[t]
    \centering
    \caption{Word error rate (WER) [\%] on LibriSpeech-\{100h, 960h\} and TEDLIUM2. Output subword vocabulary size was set to 16k for LibriSpeech-100h and TEDLIUM2, and 32k for LibriSpeech-960h. We did not use language model or beam-search during decoding.}
    \label{tb:main_results}
    \renewcommand{\arraystretch}{0.95}
    \scalebox{0.95}{
    \begin{tabular}{llcccccccccc}
        \toprule
        \multicolumn{2}{l}{\multirow{2}{*}[-8pt]{\textbf{Model}}} & \multicolumn{4}{c}{\textbf{LibriSpeech-100h}} & \multicolumn{4}{c}{\textbf{LibriSpeech-960h}} & \multicolumn{2}{c}{\textbf{TEDLIUM2}} \\
        \cmidrule(l{0.3em}r{0.3em}){3-6}\cmidrule(l{0.3em}r{0.3em}){7-10}\cmidrule(l{0.3em}r{0.3em}){11-12}
        & & \multicolumn{2}{c}{Dev WER} & \multicolumn{2}{c}{Test WER} & \multicolumn{2}{c}{Dev WER} & \multicolumn{2}{c}{Test WER} & \multirow{2}{*}[0pt]{Dev WER} & \multirow{2}{*}[0pt]{Test WER} \\
        & & clean & other & clean & other & clean & other & clean & other \\
        \midrule
        \multirow{4}{*}[0pt]{Transformer} & CTC & 11.5 & 24.8 & 11.8 & 25.5 & \phantom{0}4.2 & 10.0 & \phantom{0}4.5 & \phantom{0}9.9 & 11.8 & 10.7 \\
        & SC-CTC & \phantom{0}8.9 & 21.0 & \phantom{0}9.1 & 21.7 & \phantom{0}3.2 & \phantom{0}8.2 & \phantom{0}3.5 & \phantom{0}8.2 & \phantom{0}9.4 & \textbf{\phantom{0}8.6} \\
        & HC-CTC & \textbf{\phantom{0}8.2} & \textbf{19.9} & \textbf{\phantom{0}8.4} & \textbf{20.6} & \textbf{\phantom{0}3.1} & \textbf{\phantom{0}8.0} & \textbf{\phantom{0}3.4} & \textbf{\phantom{0}8.0} & \textbf{\phantom{0}9.1} & \textbf{\phantom{0}8.6} \\
        & ParaCTC & 10.4 & 24.0 & 10.9 & 24.3 & \phantom{0}4.6 & 10.3 & \phantom{0}4.8 & 10.3 & 10.9 & 10.2 \\
        \midrule
        \multirow{2}{*}[0pt]{Conformer} & SC-CTC & \phantom{0}7.1 & 17.7 & \phantom{0}7.7 & 18.3 & \textbf{\phantom{0}2.8} & \textbf{\phantom{0}6.7} & \textbf{\phantom{0}3.0} & \phantom{0}6.9 & \phantom{0}8.5 & \phantom{0}7.8 \\
        & HC-CTC & \textbf{\phantom{0}6.9} & \textbf{17.1} & \textbf{\phantom{0}7.1} & \textbf{17.8} & \textbf{\phantom{0}2.8} & \phantom{0}6.9 & \textbf{\phantom{0}3.0} & \textbf{\phantom{0}6.8} & \textbf{\phantom{0}8.0} & \textbf{\phantom{0}7.6} \\
        \bottomrule
    \end{tabular}}
    \vspace{-4mm}
\end{table*}

\vspace{-1.5mm}
\subsection{Proposed hierarchical conditional model}
\vspace{-1.5mm}
Figure~\ref{fig:proposed} represents an overview of the proposed hierarchical conditional model of end-to-end ASR.
It is similar to the intermediate CTC training, but
the granularity of subword units is gradually increased to word-level as the sequence transduction proceeds in the self-attention layers.
Let $Y^{(k)} \!=\! ( y^{(k)}_{l} \in \mathcal{V}^{(k)} | l\!=\!1,\dots,L^{(k)} )$ be an $L^{(k)}$-length target subword sequence
of the $k$-th CTC loss, 
which is generated by the corresponding subword segmenter with a vocabulary of $\mathcal{V}^{(k)}$.
We hierarchically increase the vocabulary size, as the position of the CTC loss becomes close to the output layer (i.e., $|\mathcal{V}^{(<K)}|\!<\!|\mathcal{V}^{(K)}|$).
Given the target sequences with different units,
the objective of the proposed model is defined by modifying Eq.~\eqref{eq:interctc_loss} as follows:
\begin{equation}
    \mathcal{L}_{\mathsf{hc\text{-}ctc}} = \frac{1}{K} \left\{ \mathcal{L}_{\mathsf{ctc}}(Y^{(K)} | X^{(E)}) + \sum_{k=1}^{K-1} \mathcal{L}_{\mathsf{ctc}}(Y^{(k)} | X^{(\lfloor \frac{kE}{K} \rfloor)}) \right\}.
    \label{eq:hierctc_loss}
\end{equation}
If the vocabulary size of each target sequence is the same,
Eq.~\eqref{eq:hierctc_loss} is equal to Eq.~\eqref{eq:interctc_loss}.
With the conditioning mechanism realized by Eq.~\eqref{eq:tf_enc3},
each CTC loss calculation in Eq.~\eqref{eq:hierctc_loss} is conditioned on the previously predicted sequences with lower levels of subword units:
\begin{equation}
    \label{eq:condition}
    \mathcal{L}_{\mathsf{ctc}} (Y^{(k)} | X^{(k)}) = - \log P_{\mathsf{ctc}}(Y^{(k)}|\hat{Y}^{(1)},...,\hat{Y}^{(k - 1)}, X^{(k)}),
\end{equation}
where $\hat{Y}^{(k)}$ denotes a sequence predicted by the $k$-th CTC,
which is implicitly represented by the posterior distributions of latent tokens.

In the proposed hierarchical conditional model,
we break down the word-level recognition into a process of progressively integrating subwords in a fine-to-coarse manner.
By making the shallower layers predict frequent subwords with small units and
the deeper layers predict sparse subwords with large units,
we expect the model to use a hierarchy of linguistic structures and
yield word-level representations effectively.



\vspace{-1.5mm}
\subsection{Applying CTC losses in parallel}
\vspace{-1.5mm}
To verify the effectiveness of the proposed model with the hierarchical structure,
we also consider training a model with CTC losses applied in parallel to the final layer,
which has been shown effective in several studies~\cite{li2017acoustic,sanabria2018hierarchical,kremer2018inductive,heba19char}.
The objective for the parallel CTC losses is defined by modifying
Eq.~\eqref{eq:hierctc_loss} as
\begin{equation}
    \mathcal{L}_{\mathsf{paractc}} = \frac{1}{K} \left\{ \mathcal{L}_{\mathsf{ctc}}(Y^{(K)} | X^{(E)}) + \sum_{k=1}^{K-1} \mathcal{L}_{\mathsf{ctc}}(Y^{(k)} | X^{(E)}) \right\}.
    \label{eq:multctc_loss}
\end{equation}
We apply a single linear layer to $X^{(E)}$ for adapting features to each CTC loss with a different granularity of subword units.

The parallel CTC training treats the predictions of multi-granular sequences equally,
where finer subword predictions provide an inductive bias to promote coarse word-level modeling~\cite{kremer2018inductive}.


\vspace{-2.4mm}
\section{Relationship to Prior Work}
\vspace{-2.2mm}
\label{sec:related_works}
Several studies have explored introducing auxiliary CTC losses to intermediate model layers and 
demonstrated its effectiveness for improving various end-to-end ASR systems, based on
attention-based sequence-to-sequence~\cite{kim2017joint,moriya2018multi},
recurrent neural network transducer~\cite{jeon2021multitask}, and
CTC~\cite{zweig2017advances,tjandra2020deja,chi2020align,lee2021intermediate}.
For the CTC-based system,
hierarchically applying low-level supervision (e.g., phonemes) to the intermediate CTC losses has shown to improve a primary CTC loss with higher-level recognition~\cite{fernandez2007sequence,toshniwal2017multitask,rao2017multi,krishna2018hierarchical,sanabria2018hierarchical}.
The proposed model can be considered an extension of these hierarchical CTC-based models.
However,
our work differs from prior work in the following perspectives.
1) Each CTC loss is explicitly conditioned on the sequences predicted previously at lower abstraction levels.
We expect the model to maintain subwords that contribute to composing a word-level sequence and promote the CTC training with conditional dependencies~\cite{nozaki2021relaxing}.
2) Given that, in recent studies~\cite{tjandra2020deja,lee2021intermediate}, the intermediate CTC losses are effective even without the hierarchical supervision,
we carefully conduct a comparative experiment and further analyze the effectiveness of hierarchical modeling.
3) We only use subwords for target sequences,
which does not require additional labeling effort and is easy to control the granularity of target sequences.
4) We evaluate models using the recent state-of-the-art architectures (i.e., Transformer~\cite{vaswani2017attention} and Conformer~\cite{gulati2020conformer}).

\vspace{-2.4mm}
\section{Experiments}
\vspace{-2.2mm}
\label{sec:experiments}

\subsection{Experimental setup}
\vspace{-1mm}
\noindent\textbf{Data:}
The experiments were carried out using the LibriSpeech (LS)~\cite{panayotov2015librispeech} and TEDLIUM2 (TED2)~\cite{rousseau2014enhancing} datasets.
LS consists of utterances from read English audio books.
We trained the models using the 100-hour subset (LS-100) or the 960-hour full set (LS-960).
TED2 consists of utterances from English Ted Talks and contains 210 hours of training data.
For each dataset, 
we used the standard development and test sets.
As input speech features, 
we extracted 80 mel-scale filterbank coefficients with three-dimensional pitch features using Kaldi~\cite{povey2011kaldi},
which were augmented by speed perturbation and SpecAugment~\cite{park2019specaugment}.
We used SentencePiece~\cite{kudo2018subword} to construct subword vocabularies for each dataset.

\noindent\textbf{Evaluated models:}
\textbf{CTC} denotes a standard CTC-based model trained with $\mathcal{L}_{\mathsf{ctc}}$ from Eq.~\eqref{eq:ctc_loss}~\cite{graves2014towards}.
\textbf{SC-CTC} is a conventional model trained with the intermediate CTC losses~\cite{tjandra2020deja,lee2021intermediate} and the self-conditioning mechanism~\cite{nozaki2021relaxing} defined by $\mathcal{L}_{\mathsf{sc\text{-}ctc}}$ in Eq.~\eqref{eq:interctc_loss}.
\textbf{HC-CTC} is the proposed hierarchical conditional model trained with $\mathcal{L}_{\mathsf{hc\text{-}ctc}}$ from Eq.~\eqref{eq:hierctc_loss}.
\textbf{ParaCTC} is a conventional model trained with the parallel CTC losses defined by $\mathcal{L}_{\mathsf{paractc}}$ in Eq.~\eqref{eq:multctc_loss}~\cite{li2017acoustic,sanabria2018hierarchical,kremer2018inductive}.


\noindent\textbf{Training and decoding configurations:} All experiments were conducted using ESPnet~\cite{watanabe2018espnet}.
We used the Transformer~\cite{vaswani2017attention} architecture to train the above models,
which consisted of two CNN layers followed by a stack of 18 self-attention layers.
The number of heads $d_{\mathsf{h}}$, dimension of a self-attention layer $d_{\mathsf{model}}$,
and dimension of a feed-forward network $d_{\mathsf{ff}}$ were set to 4, 256, and 2048, respectively.
We also trained the models using the Conformer architecture~\cite{gulati2020conformer},
which had a kernel size of 15 and the same configurations as the Transformer-based models, except $d_{\mathsf{ff}}$ was set to 1024.
The models were trained up to 100 epochs.
For models with multiple CTC losses (i.e., SC-CTC, HC-CTC, and ParaCTC),
we set the total number of losses to 3 ($K=3$).
The output vocabulary sizes for LS-100, LS-960, and TED2 were set to 16384, 32768, and 16384, respectively.
Each vocabulary size was determined on the basis of the maximum number we could set using SentencePiece,
which is large enough to be considered as word-level.
SC-CTC had intermediate losses with the same vocabulary size as the output's.
For HC-CTC and ParaCTC,
we set ($|\mathcal{V}^{(1)}|$, $|\mathcal{V}^{(2)}|$, $|\mathcal{V}^{(3)}|$) to (256, 2048, 16384) for LS-100 and TED2, and 
(512, 4096, 32768) for LS-960.
After training, a final model was obtained by averaging model parameters over 10 to 20 checkpoints with the best validation performance.
During decoding,
we did not use any language model and
carried out the best path decoding of CTC~\cite{graves2006connectionist}.
Our implementations are publicly available to ensure reproducibility (see Sec.~\ref{sec:intro}).

\vspace{-2mm}
\subsection{Main results}
\vspace{-1mm}
Table~\ref{tb:main_results} lists the results on LS-100, LS-960, and TED2 in terms of the word error rate (WER).
Looking at the Transformer results,
all the models trained with multiple CTC losses led to an improvement over the standard CTC-based model.
Especially, SC-CTC and HC-CTC significantly reduced the WER on all of the tasks.
On LS-100,
HC-CTC showed a clear improvement over SC-CTC,
indicating the effectiveness of hierarchically increasing subword units.
In contrast,
on LS-960 and TED2 with more data,
the performance gap was reduced, and
HC-CTC performed slightly better than SC-CTC.
Therefore,
it can be concluded that our model is particularly effective for smaller-scale data,
where the word-level units are likely to become sparser.
SC-CTC was capable of handling word-level units when there is a sufficient amount of data.
However,
the large vocabulary-sized softmax calculation (in Eq.~\eqref{eq:tf_enc3}) led to a severe slow-down of the SC-CTC training and inference processes.
HC-CTC, on the other hand, was able to perform faster training and inference,
using finer units for the losses from intermediate layers.
Due to the same reason regarding the softmax calculation, the model size of HC-CTC was much smaller than that of SC-CTC (e.g., 36.4M vs. 67.6M on LS-960).
By comparing HC-CTC with ParaCTC,
HC-CTC achieved much lower WERs on all tasks,
demonstrating the effectiveness of applying CTC losses to intermediate layers as well as
gradually increasing the subword units in a hierarchical manner.

Using Conformer further improved the performance of SC-CTC and HC-CTC, and
HC-CTC again achieved more favorable performance than SC-CTC with faster training and inference.
Our Conformer results are comparable with other strong CTC-based models of the same size~\cite{ng2021pushing,majumdar2021citrinet,higuchi2021comparative},
even without exhaustive tuning.

\vspace{-2.5mm}
\subsection{Analysis on subword vocabulary size}
\vspace{-1.2mm}
\label{ssec:analysis_subwords}
\begin{table}[t]
    \centering
    \caption{WER [\%] on LS-100 dev.\ sets for Transformer-based models trained with different combinations of subword vocabulary sizes.}
    \label{tb:analysis_subwords}
    \renewcommand{\arraystretch}{0.95}
    \scalebox{0.95}{
    \begin{tabular}{lcccc}
        \toprule
        \textbf{Model} & $|\mathcal{V}^{(1)}|$-$|\mathcal{V}^{(2)}|$-$|\mathcal{V}^{(3)}|$ & dev-clean & dev-other \\
        \midrule
        SC-CTC & \phantom{/}256\phantom{/}-\phantom{/}256\phantom{/}-\phantom{/}256\phantom{/} & 8.4 & 22.8 \\
        SC-CTC & \phantom{/0}2k\phantom{/}-\phantom{/0}2k\phantom{/}-\phantom{/0}2k\phantom{/} & 8.5 & 22.0 \\
        SC-CTC & \phantom{/}16k\phantom{/}-\phantom{/}16k\phantom{/}-\phantom{/}16k\phantom{/} & 8.9 & 21.0 \\
        HC-CTC & \phantom{/}256\phantom{/}-\phantom{/}256\phantom{/}-\phantom{/}16k\phantom{/} & \textbf{8.2} & 20.2 \\
        HC-CTC & \phantom{/0}2k\phantom{/}-\phantom{/0}2k\phantom{/}-\phantom{/}16k\phantom{/} & 8.4 & 20.2 \\
        \cdashlinelr{1-4}
        HC-CTC & \phantom{/}256\phantom{/}-\phantom{/0}2k\phantom{/}-\phantom{/}16k\phantom{/} & \textbf{8.2} & \textbf{19.9} \\
        \bottomrule
    \end{tabular}}
    \vspace{-4mm}
\end{table}
While using sparse word-level units can make training of an ASR model challenging~\cite{soltau2016neural},
we observed that the standard CTC-based model, with the Transformer-based architecture, benefits from training with a large subword vocabulary size.
By increasing the output vocabulary size from 256 to 16k,
the WERs for dev.\ sets changed from 11.1/28.1\% to 11.5/24.8\% on LS-360, and 
12.3\% to 11.8\% on TED2.
Similarly,
the performance on LS-960 changed from 4.6/12.1\% to 4.4/10.5\% by changing the vocabulary size from 2k to 32k.
These decent improvements from increasing the subword vocabulary size can be attributed to
compensating for the CTC's incapability of modeling output dependencies (cf. Eq.\eqref{eq:p_ctc}).

Considering the above observation,
we evaluated SC-CTC and HC-CTC with different combinations of vocabulary sizes,
focusing on Transformer-based models trained on LS-100.
From the results for SC-CTC in Table~\ref{tb:analysis_subwords},
the performance on the dev-other set improved by increasing the vocabulary size,
benefiting from the CTC training with large subword units.
HC-CTC performed better than the 16k result of SC-CTC,
indicating HC-CTC was more effective at modeling word-level recognition besides the advantage of CTC training with a large vocabulary size.
While the SC-CTC performance on the dev-clean set degraded by increasing the vocabulary size,
HC-CTC succeeded in learning robust word-level representations and achieved the lowest WER with the 16k-vocabulary size.
Comparing the HC-CTC results,
hierarchically increasing the subword units resulted in better performance than using the same vocabulary size for intermediate losses,
suggesting the importance of gradually increasing the abstraction level for
learning word-level representations effectively.

\vspace{-2.5mm}
\subsection{Importance of conditioning}
\vspace{-1.2mm}
\label{ssec:analysis_conditioning}
We studied the effectiveness of the conditioning mechanism,
which is one of the important components of the proposed model (cf. Eq.~\eqref{eq:condition}).
The Transformer-based HC-CTC was trained on LS-100 without conditioning each CTC loss (i.e., Eqs.~\eqref{eq:tf_enc1} and~\eqref{eq:tf_enc2} were used for all the intermediate layers).
Note that this model is similar to those from previous studies~\cite{fernandez2007sequence,toshniwal2017multitask,rao2017multi,krishna2018hierarchical,sanabria2018hierarchical}.
Without the conditioning mechanism,
HC-CTC achieved WERs of 8.7/20.7\% and 9.0/21.3\% on dev.\ sets and test sets, respectively.
While these results are better than those obtained from CTC, SC-CTC, and ParaCTC in Table~\ref{tb:main_results},
HC-CTC with the conditioning mechanism achieved much lower WERs.
Overall,
we can conclude that 1) hierarchical modeling based on multi-granular subword units
as well as 2) the conditioning mechanism for explicitly maintaining lower levels of predictions
are effective for learning word-level representations.


\vspace{-2.5mm}
\subsection{Attention visualization}
\vspace{-1.2mm}
\label{ssec:analysis_attention}
\begin{figure}[t]
    \centering
    \begin{tabular}{c}
        \rotatebox[origin=c]{90}{\textbf{(a) CTC}}\hspace{1mm}
        \begin{minipage}{0.92\columnwidth}
            \centering
            \includegraphics[width=\linewidth]{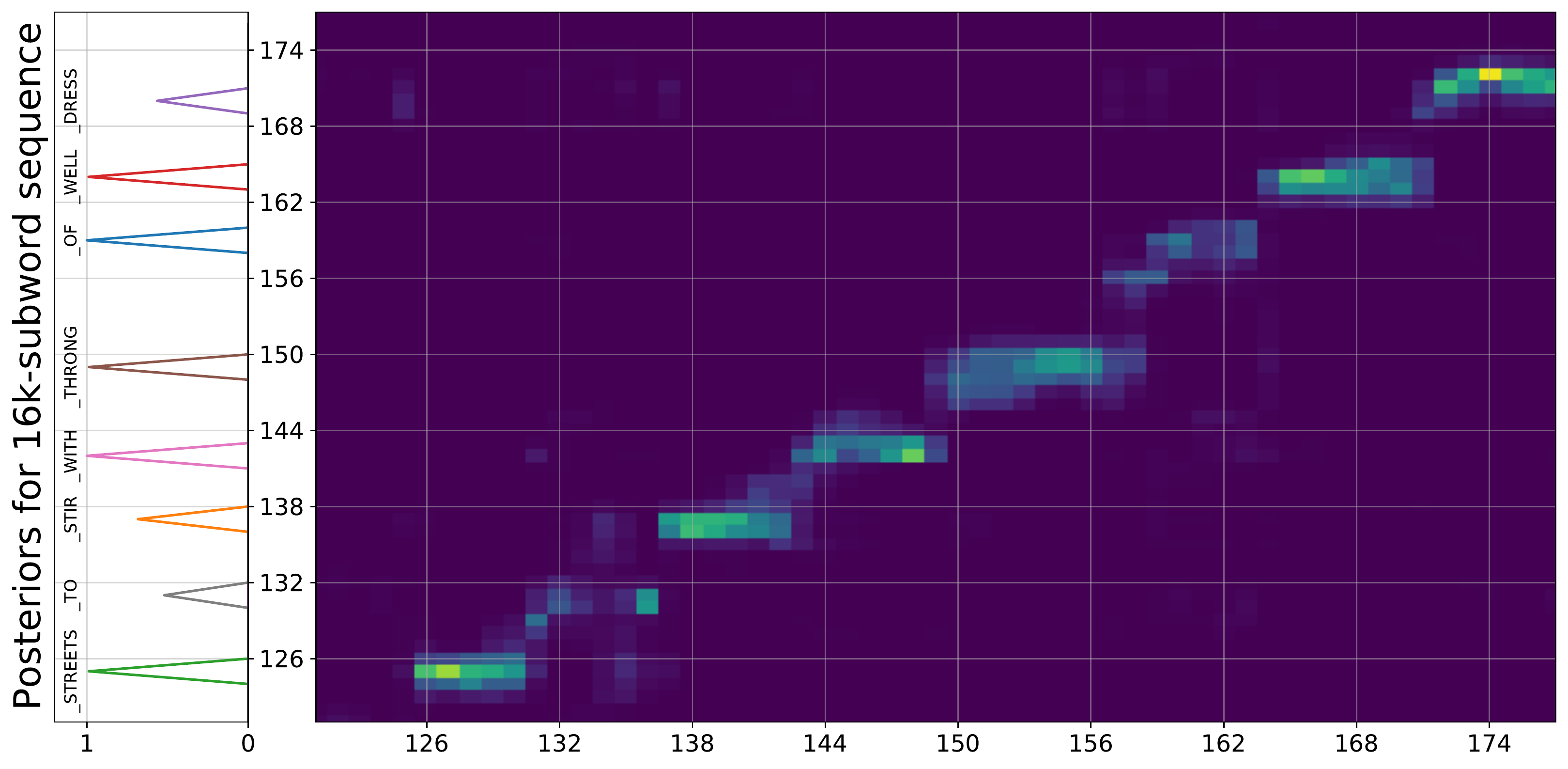}
        \end{minipage}\\
        \rotatebox[origin=c]{90}{\textbf{(b) HC-CTC}}\hspace{1mm}
        \begin{minipage}{0.92\columnwidth}
            \centering
            \vspace{1.2mm}
            \includegraphics[width=\linewidth]{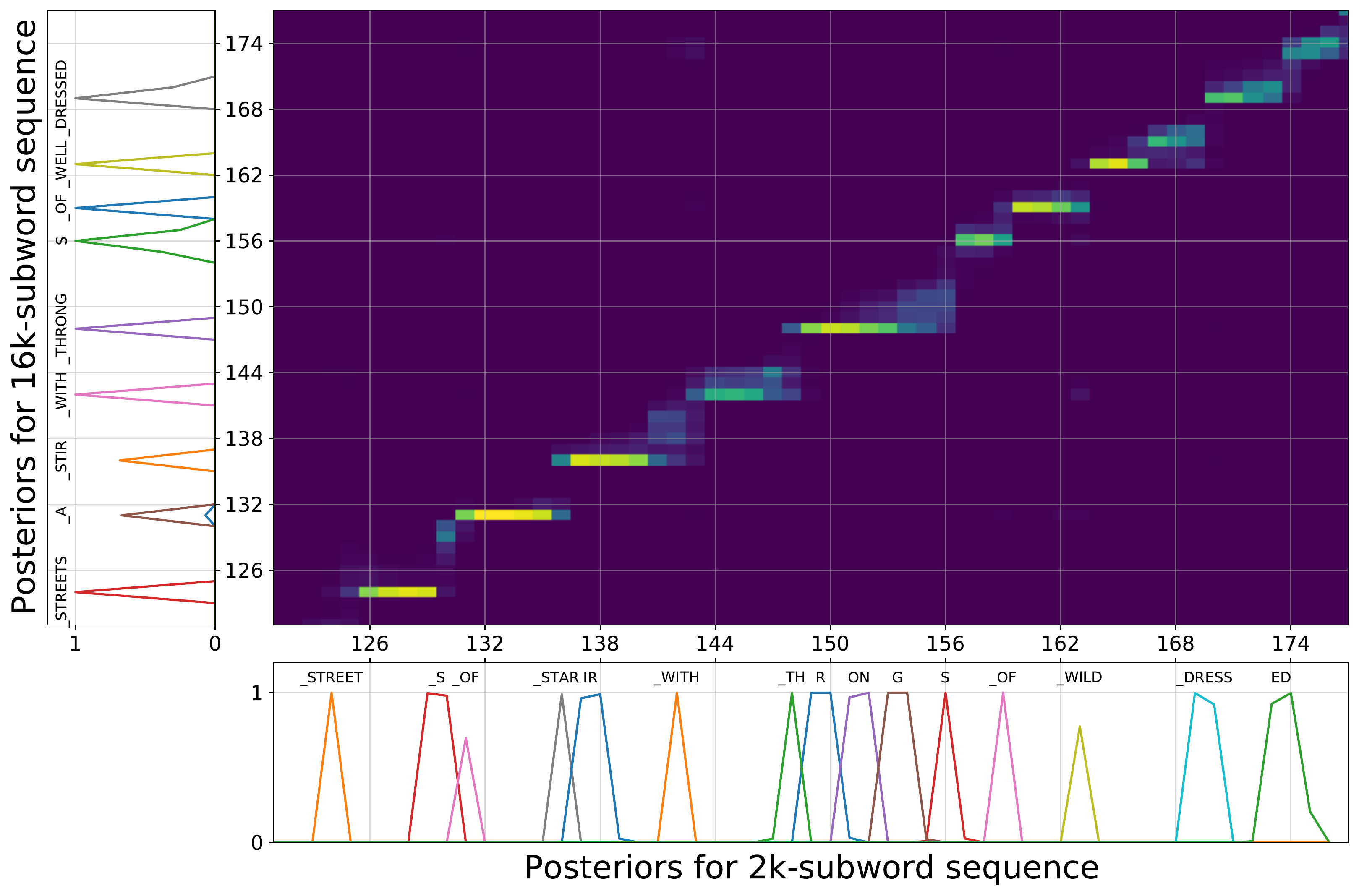}
        \end{minipage}
    \end{tabular}
    \vspace{-3mm}
    \caption{Attention visualization of (a) CTC and (b) HC-CTC trained on LS-100 from Table~\ref{tb:main_results}. We manually chose partial utterance from dev-other set (116-288045-0000), transcription of which is \textit{``STREETS ASTIR WITH THRONGS OF WELL DRESSED''}.}
    \label{fig:attention}
    \vspace{-4mm}
\end{figure}
Figure~\ref{fig:attention} visualizes attention weights between a source (x-axis) and target (y-axis) sequences,
comparing Transformer-based (a) CTC and (b) HC-CTC trained on LS-100 from Table~\ref{tb:main_results}.
We focused on weights that seemed to contribute to predicting a 16k-subword sequence in the final CTC (from the 18-th layer).
For HC-CTC, we show the CTC posteriors (from the 12-th layer) for predicting a 2k-subword sequence in advance to see the relationship to the 16k prediction.
Comparing the overall weights,
HC-CTC learned more solid and confident weights than CTC.
HC-CTC seemed to exploit the lower-level 2k predictions to detect important frames for predicting each token,
effectively composing complex word-level tokens using the lower-level tokens.
For example, HC-CTC successfully recognized the words ``THRONGS'' and ``DRESSED'' with proper conjunctions,
while CTC failed to handle these infrequent words.

\vspace{-2mm}
\section{Conclusions}
\vspace{-2mm}
\label{sec:conclusions}
We proposed a hierarchical conditional model of CTC-based end-to-end ASR.
We trained the model by gradually increasing the subword units for CTC losses applied to intermediate layers.
Each CTC loss was conditioned on the sequences with lower abstraction
to compose higher-level prediction.
Experimental results and in-depth analysis demonstrated that our model effectively learned word-level representations for improving ASR performance.
Future work includes introducing an additional decoder network~\cite{higuchi2020mask} and
using acoustic-based subword unit for lower-level predictions~\cite{xu2019improving,zhou2021acoustic}.

\vspace{-2mm}
\section{Acknowledgement}
\vspace{-2mm}
This work was supported in part by JST ACT-X (JPMJAX210J).

\newpage

\fontsize{8.7pt}{0pt}\selectfont
\bibliographystyle{IEEEbib}
\bibliography{refs}

\end{document}